\documentclass [epj,final] {svjour} 

\usepackage{epsfig,graphicx,tabularx}
\usepackage{amsmath,amssymb}
\usepackage{color,braket}

\newcommand{\mb}{\mathbf}
\newcommand{\be}{\begin{equation}}
\newcommand{\ee}{\end{equation}}

\begin{document}

\title{A study of coherently coupled two-component Bose-Einstein condensates}

\authorrunning{M. Abad and A. Recati}
\titlerunning{A study of coherently coupled two-component Bose-Einstein condensates}

\author{Marta Abad \and Alessio Recati}
\institute{INO-CNR BEC Center and Dipartimento di Fisica, Universit\`a di Trento, 38123 Povo, Italy}

\abstract{
We present a self-consistent study of coherently coupled two-component Bose-Einstein condensates. 
Finite spin-flipping coupling changes the first order demixing phase transition for Bose-Bose mixtures to a second order phase transition between an unpolarized and a polarized state. We analise the excitation spectrum and the structure factor along the transition for a homogeneous system. We discuss the main differences at the transition between a coherent coupled gas and a two-component mixture. 
We finally study the ground state when spin-(in)dependent trapping potentials are added to the system, focusing on optical lattices, which give rise to interesting new configurations. }


\maketitle

\section{Introduction}

The physics of multi-component condensates is very rich due to the possibility of vector order parameters and the presence of different zero-temperature phases.

In the cold gases context the ability to tune and to engineer single- and two-body properties permits many implementations of such a system. This toolbox allows to address very different and interesting phenomena like Andreev-Bashkin effect~\cite{andreev}, persistent currents~\cite{ZoranSC}, (internal) Josephson effect~\cite{ChapmanJJ,Zibold2010}, Schr\"odinger-cat- and twin-Fock-like states~\cite{Cirac1998,TwinExp}, analogues of quantum gravity~\cite{Uwe,Liberati2006,Sindoni2011}, spin textures~\cite{texture}, or the most recent and fashionable field of non-abelian gauges~\cite{DalibardRMP}, just to cite a few of them.

In the present work we consider one of the easiest implementations, namely a 2-component (spinor) condensate with an external field that drives the population transfer (spin-flipping) between the two atomic levels. 
In spite of the apparent simplicity of the problem, the physics it contains is very rich, which is reflected in the vast amounts of literature generated in the past decades.
In this article we aim at providing an understanding of the most fundamental features arising from the coupling. 
Many properties of such a system have been already understood and addressed in the literature. For the purposes of this article, the most relevant works refer to the structure of the ground state and its excitation spectrum~\cite{Goldstein1997,Search2001,Tommasini2003,Lee2004,Liang2010}.



Two component condensates are interesting because they constitute the generalization of the well known Rabi problem of atom optics to interacting extended non-linear systems (see, e.g., one of the first experiments \cite{Cornell99}) and many properties can be indeed understood using a Bloch sphere representation. Under some circumstances it has been shown that two component spinors allow for a description in terms of a dressed state basis \cite{Blakie1999,Search2001b,Jenkins2003,Nicklas2011}.  Moreover the population transfer between the two levels turns out to be described by Josephson dynamics, leading to what is known as internal Josephson effects (see, e.g., \cite{Leggett2001}). 

With the increase of atomic species which can be condensed, spinor systems offer new interesting possibilities, for instance when species-dependent external potentials can be created (see, e.g., the proposals in~\cite{Recati2003,Daley2008}).
The relation between the latter system and the condensates with artificial spin-orbit-like interactions, which is clearly built on the physics of two-component spinor condensates, would be useful for the field.

For these reasons in the following we try to give a self-consistent and complete description of the system collecting together a number of results.
Some of the results we obtain are scattered in literature and proper references are provided. 
New results on the dynamics of these systems and on their behaviour in the presence of external potentials are presented.

The structure of the article will be as follows. In Sec.~\ref{SecGS} we analyse the ground state of the homogeneous system, which shows a phase transition between a neutral (GS1) and a polarised state (GS2). 
The static compressibility and susceptibility are also studied. The latter has a divergence at the symmetry breaking point.

The elementary excitations of the two component system are addressed in Sec.~\ref{SecExc}. First we write an effective quantum hydrodynamic theory to get an insight into the structure of modes, Sec.~\ref{SecHydro}. The elementary excitations are then studied in detail within a Bogoliubov approach in Sec.~\ref{SecBogExc}.

The knowledge of the the spectrum allows us to calculate in Sec.~\ref{SecSFacGS1} the density and spin structure factors. Such quantities can be measured experimentally also in the trapped case, being related to the local fluctuations of the density and of the polarisation.

The case of trapped gases is discussed in Sec.~\ref{SecTrap}. First we address the case where the potentials acting on the two species are the same, Sec.~\ref{SecHTrap}. In Sec.~\ref{SecOL} we concentrate on the case of an optical lattice acting only on one component. 

\section{Ground state of homogeneous spinor condensates}\label{SecGS}

We consider a homogeneous spinor condensate whose two components $a$ and $b$ interact both via $s$-wave contact interactions and via a coherent coupling. Within the mean-field framework the system is described by coupled Gross-Pitaevskii equations for the spinor components $\Psi_a(\mathbf{r},t)$ and $\Psi_b(\mathbf{r},t)$
\begin{align}
 i\hbar\frac{\partial }{\partial t}\Psi_a=&\left[ -\frac{\hbar^2\nabla^2}{2m} +V_a + g_{a}|\Psi_a|^2 + g_{ab}|\Psi_b|^2  \right]\Psi_a +\nonumber\\
&+  \Omega\Psi_b \label{tdgppsia} \\
 i\hbar\frac{\partial }{\partial t}\Psi_b=&\left[ -\frac{\hbar^2\nabla^2}{2m} +V_b +  g_{b}|\Psi_b|^2 + g_{ab}|\Psi_a|^2  \right]\Psi_b +\nonumber\\
&+  \Omega^*\Psi_a \label{tdgppsib}
\end{align}
where $m$ is the atomic mass.  
The contact interaction coupling constants are given by $g_i=4\pi\hbar^2 a_i/m$, with $i=a,b,ab$, where $a_a$ and $a_b$ are the $s$-wave scattering lengths for components $a$ and $b$, and $a_{ab}$ that associated to the interaction between $a$ and $b$. 
The term $\Omega$ introduces a coherent coupling between the two components, which gives rise to phase correlations between the two fluids, in contrast to the density-density correlations coming from the interspecies interaction $g_{ab}$. 
Depending on the physical system, this term can have its origin on either a two-photon (Raman) process  or a direct Rabi coupling between the components. 
For the homogeneous system, the external potentials are $V_a=V_b=0$, which is the situation we consider in this article except in Sec.~\ref{SecTrap}.

Due to the flipping term only the total number of particles (total density in the uniform system) $n=n_a+n_b$ is conserved. Thus the chemical potential $\mu$ is the same for both components and the stationary states evolve as 
\begin{equation}
 \Psi_\sigma(t)=e^{-i\mu t/\hbar}\psi_\sigma \;\;\; \sigma=a,b.
\end{equation}
It is convenient to write the spinor components in terms of the density $n_\sigma$ and the phase $\phi_\sigma$ 
\begin{equation}
 \psi_\sigma=\sqrt{n_\sigma}e^{i\phi_\sigma}\ .\label{gswf}
\end{equation}

The ground state of the system has been described in the literature~\cite{Tommasini2003,Lee2004,Blakie1999,Search2001b}, but here we revisit it introducing a new convenient notation. The ground state is given by the values of densities and phases which minimize the energy per unit volume
\begin{align}
   e(n_a,n_b) =&\frac{1}{2}g_an_a^2+\frac{1}{2}g_bn_b^2 + g_{ab}n_an_b + \nonumber\\ 
    & + 2|\Omega|\cos\phi\sqrt{n_an_b} -\mu (n_a+n_b) \label{ecan}
\end{align}
where we have introduced the phase $\phi\equiv\phi_{ba}+\phi_\Omega$, in terms of the phase difference $\phi_{ba}=\phi_b-\phi_a$ and the phase of the Rabi coupling, given by $\Omega=|\Omega|e^{i\phi_\Omega}$. The configuration with minimum energy corresponds to $\cos\phi=-1$. For $\Omega$ real ($\phi_\Omega=0,\pi$) this means $\phi_{ba}=\pi$ for $\Omega>0$ and $\phi_{ba}=0$ for $\Omega<0$; for $\Omega$ complex, the equilibrium value of $\phi_{ba}$ is such that it satisfies $\phi_{ba}+\phi_\Omega=(2n+1)\pi$ with $n\in Z$. Notice that the condition $\cos\phi=+1$ can give rise to an extremum of the energy \cite{Goldstein1997,Search2001,Tommasini2003}, but it will never be the global minimum (in fact it is a saddle-point in the energy landscape). 

\begin{figure}\centering
   \epsfig{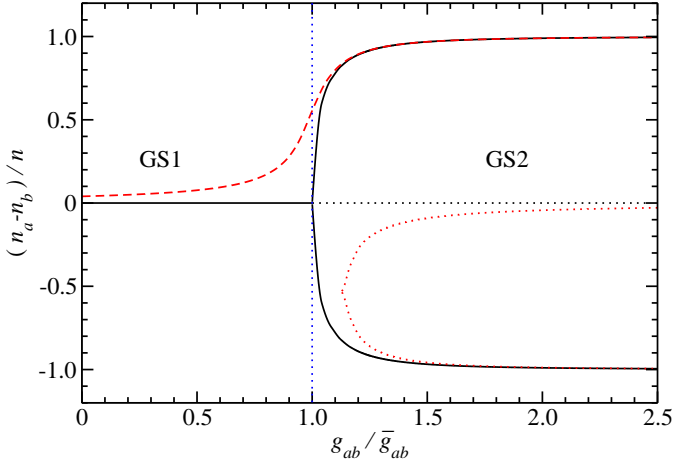}
  \caption{Different ground states (GS1 and GS2) exhibited by the two-component spinor system as a function of $g_{ab}/\bar{g}_{ab}$. In solid: $g_a=g_b$; In dashed: $\delta g=0.1g$ (see text). In dotted: unstable solutions. In all cases $\Omega=0.1\, gn$.}
  \label{figgs1}
\end{figure}

The equilibrium configuration is then characterised by the density difference $n_a-n_b$. 
The structure of the ground state is better understood in the symmetric case $g_a=g_b\equiv g$, when the equilibrium solutions must satisfy the equation
\begin{equation}
 \left(g-g_{ab}+\frac{|\Omega|}{\sqrt{n_an_b}}\right)(n_a-n_b)=0
\end{equation}
This equation admits the following solutions \begin{align}
 &\text{(GS1)}\quad n_a-n_b=0; \label{nsym} \\
 &\text{(GS2)}\quad (n_a-n_b)_{\pm}=\pm n\sqrt{1 - \left(\frac{2|\Omega|}{(g-g_{ab})n}\right)^2}, \label{npm}
\end{align}
corresponding to neutral (GS1) and polarised (GS2) ground states.
Introducing the parameter $\bar{g}_{ab}=g+2\Omega/n$ one finds that GS1 (GS2) has the minimum energy provided $g_{ab}<\bar{g}_{ab}$ ($g_{ab}>\bar{g}_{ab}$). In Fig.~\ref{figgs1} we report the polarisation of the condensate as a function of the inter-species interaction $g_{ab}$. At the critical point $g_{ab}=\bar{g}_{ab}$ there exists a bifurcation in the ground state solutions, which hints at a second-order phase transition.
This bifurcation has been measured in the experiment reported in Ref.~\cite{Zibold2010} and has been predicted to be related to the emergence of a Schr\"odinger cat state \cite{Cirac1998}.
It is worth remembering that in a 2-state condensate without the flipping term (i.e., $n_a$, $n_b$ conserved) the condition $g_{ab}=g$ distinguishes a homogeneous from a phase separated state (see, e.g., \cite{BECPhaseSep}). Clearly the presence of the coherent coupling shifts the critical value to higher values of interspecies interaction, and prevents phase separation by instead creating a polarisation.

Let us briefly  comment here on the relation between the scenario above and the internal Josephson effect (see also Ref.~\cite{Lee2004}), which is usually addressed from Eqs.~(\ref{tdgppsia}) and (\ref{tdgppsib}), recognizing $\Omega$ as the weak coupling. The phase we have named GS1 corresponds to a fixed point in the Josephson Hamiltonian around which closed orbits exist, with vanishing mean polarisation (or population imbalance) and a phase difference around $\pi$, giving rise to plasma-like oscillations. Instead, around the fixed point GS2 the polarisation is finite, leading to self-trapping dynamics (not a running phase-mode, though, since the mean phase difference is locked to $\pi$).

\begin{figure}[h]\centering
 \epsfig{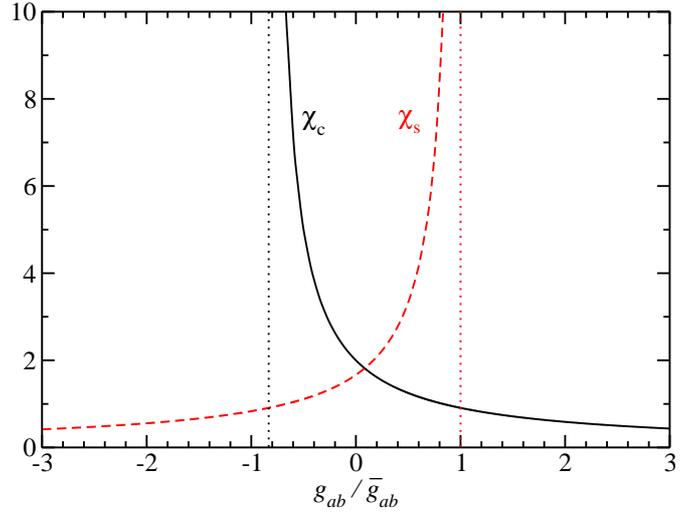}
 \caption{Density (solid) and spin (dashed) static response functions as a function of the coupling constant of the interspecies contact interaction, $g_{ab}$, above GS1, for $\Omega=0.1\,gn$.}\label{figresponse}
\end{figure}

The existence of the phase transition can be more clearly identified looking at the susceptibility of GS1, which within linear response is given by
\begin{equation}
  \chi_s=\lim_{\eta\to0}\frac{n_a - n_b}{\eta}=\frac{2}{g-g_{ab}+2|\Omega|/n} \,,
\end{equation}
and diverges at the phase transition.
The compressibility on the other hand reads
\begin{equation}
  \chi_d=\lim_{\eta\to0}\frac{ n_a +  n_b}{\eta}=\frac{2}{g+g_{ab}}\,. 
\end{equation}
It diverges for $g\to -g_{ab}$, indicating collapse of the condensate.
The quantity $\eta$ is the amplitude of a small perturbation of the energy, characterising the term $-\eta(n_a\pm n_b)$ added to Eq.~(\ref{ecan}), for a density perturbation or a spin perturbation, respectively. 
The susceptibility and compressibility are plotted in Fig.~\ref{figresponse} as a function of $g_{ab}$ for a system in the symmetric GS1.

Introducing an asymmetry in the interaction potential, i.e., $g_a \neq g_b$ the ground state always shows a certain polarisation and the degeneracy (bifurcation) is removed. We take in particular $g_a=g$ and $g_b=g+\delta g>g$.
For GS1 and $g_{ab}\ll \bar{g}_{ab}$ a small $\delta g$ creates a linear polarisation in the system given by
\begin{equation}
 \delta n =n_a-n_b = \frac{n}{2}\frac{\delta g}{g-g_{ab}+2|\Omega|/n+\delta g}. \label{deltan}
\end{equation}
Obviously for $\delta g>0$ we always have $n_a-n_b >0$ (see dashed line in Fig.~\ref{figgs1}). 
As $g_{ab}\to \bar{g}_{ab}$ even a small $\delta g$ strongly polarises the system, and for $g_{ab}>\bar{g}_{ab}$ the polarisation fastly saturates to the value of GS2 for $\delta g=0$. On the other hand the lower branch becomes a high energy minimum and the $n_b=n_a$ branch becomes a maximum.

\section{Excitation spectrum}\label{SecExc}

\subsection{Hydrodynamic picture: hybridization of spin and density modes}\label{SecHydro}

In the present Section we use a perturbative hydrodynamic approach in order to get an insight into the nature of the excitation spectrum. In the next Section a Bogoliubov approach is employed for a more quantitative analysis. Let us introduce the fluctuation fields $\Pi_\sigma$ and $\phi_\sigma$ for the densities and the phases, respectively. The  various contributions to the effective hydrodynamic energy functional for the excitations, $E_{HD}=E_0+V_{ab}$, read \cite{Popov}
\begin{align}
 E_0=&\sum_{\sigma=a,b}\int\left[\frac{\hbar^2 n_\sigma}{2m}(\nabla\phi_\sigma)^2+\frac{mc_\sigma^2 }{2 n_\sigma}\Pi_\sigma^2\right]; \label{E0hd}\\
V_{ab}=&\left(g_{ab}-\frac{\Omega}{2 \bar n}\right)\int\Pi_a\Pi_b-\frac{\Omega\bar n}{4}\!\int\!\left[\left(\frac{\Pi_a}{n_a}\right)^2\!\!+\left(\frac{\Pi_b}{n_b}\right)^2\right]\nonumber\\
&+\Omega\bar n\int(\phi_a-\phi_b)^2,
\end{align}
where as usual we keep the terms up to second order in the fields and we define 
$\bar n=\sqrt{n_a n_b}$ and $c_\sigma$ the speeds of sound of the two components when the coupling is switched off.

The analysis is most enlightening for the symmetric case $g_a=g_b=g$, when $c_\sigma^2=c^2=gn/m$ and $\bar n=n/2$. The energy is obviously diagonalised by the density (or in-phase), $\xi_d=(\xi_a+\xi_b)/2$, and spin-density (or out-of-phase), $\xi_s=(\xi_a-\xi_b)/\sqrt{2}$, fluctuation fields, with $\xi=\Pi,\;\phi$.
The energy functional can be written in terms of the new fields as
\begin{align}
 E_{HD}&=\int\left[\frac{\hbar^2 n}{4m}(\nabla\phi_d)^2+\frac{mc_d^2 }{n}\Pi_d^2\right]+ \nonumber\\
&+\int\left[\frac{\hbar^2 n}{4m}(\nabla\phi_s)^2+\frac{mc_s^2 }{n}\Pi_s^2 +\Omega n \phi_s^2\right], \label{Eds}
\end{align}
where we have introduced the density $c_d$ and spin $c_s$ sound speeds (see Eqs. (\ref{wdHD}) and (\ref{wsHD})). From Eq.~(\ref{Eds})  one immediately sees that while the density sector remains linear and gapless, a gap can appear in the spin sector and the two branches cross for a certain value of momentum. Indeed the equations of motion derived from the previous energy functional give the dispersion relations 
\begin{align}
&\omega_d=c_d |k|=\sqrt{\frac{(g+g_{ab})n}{2 m}}|k| \label{wdHD}\\
&\omega_s=c_s\sqrt{k^2+4 m \Omega}=\sqrt{\left(\frac{(g-g_{ab})n}{2 m}+\frac{\Omega}{m}\right)(k^2+4 m \Omega)} \label{wsHD}
\end{align}
allowing for a value $k_0$ at which $\omega_d=\omega_s$.
From the dispersion relations Eqs.~(\ref{wdHD}) and (\ref{wsHD}) it is clear that such an unpolarised phase can be stable against collapse only if $g_{ab}>-g$ and against polarisation only if $g_{ab}<g+2\Omega/n$. The latter is precisely the condition we found in the previous Section for GS1 to be the lowest energy state. 
Notice that since the leading term at small momentum for the spin spectrum is $k^2$ its coefficient is wrong within hydrodynamics. To get it right it would be enough to include the next order term, i.e.  $(\nabla\Pi_s)^2/(2nm)$ \cite{Popov}, in the energy functional Eq.~(\ref{Eds}).

Any small difference between $g_a$ and $g_b$ would couple the density and spin-density leading to an hybridization of the modes and thus to an avoided crossing between the new branches.

\subsection{Bogoliubov excitations}\label{SecBogExc}

In this Section we derive the dispersion law of the two-component spinor BEC for the excitations above the different ground states within Bogoliubov approach. The dynamics of the system is given by the coupled time-dependent Gross-Pitaevskii Eqs.~(\ref{tdgppsia}) and (\ref{tdgppsib}). To address small amplitude excitations above the ground state we write the time-dependent wave functions as
\begin{align}
 \Psi_\sigma(\mb{r},t)&=e^{-i\mu t/\hbar}\left[ \psi_\sigma + e^{i\phi_\sigma}\delta \Psi_\sigma(\mathbf{r},t)  \right] \label{dpsia}
\end{align}
with $\mu$ the chemical potential of the complete system, $\psi_\sigma$ the ground state wave functions defined in Eq.~(\ref{gswf}) and $\sigma=a,b$. 
Since we are dealing with a homogeneous system we can write the excitations $\delta\Psi_\sigma$ in terms of plane waves with amplitudes $u_\sigma$ and $v_\sigma$,
\begin{align}
 \delta\Psi_\sigma&=  u_\sigma e^{i(\bf{k}\cdot\bf{r}-\omega t)}, \label{Bogclass1}\\
 \delta\Psi_\sigma^*&=  v_\sigma e^{i(\bf{k}\cdot\bf{r}-\omega t)}.\label{Bogclass2}
\end{align}
The equations for the Bogoliubov coefficients $u$'s and $v$'s can be written as
\begin{displaymath}
 \hbar\omega \left(
 \begin{array}{c}
  u_a \\ v_a\\ u_b \\v_b
 \end{array}\right)= \mathcal{L}\left(
 \begin{array}{c}
  u_a \\ v_a\\ u_b \\v_b
 \end{array}\right)
\end{displaymath}
with
\begin{equation}
 \mathcal{L}=\left(
 \begin{array}{cccc}
  h_a  & g_{a}n_a &   g_{ab}n_{ab} - |\Omega| & g_{ab}n_{ab} \\
  -g_{a}n_a & -h_a & -g_{ab}n_{ab} & - g_{ab}n_{ab} + |\Omega|\\
   g_{ab}n_{ab} - |\Omega| & g_{ab}n_{ab} & h_b & g_{b}n_b \\
  -g_{ab}n_{ab} & -g_{ab}n_{ab} + |\Omega| & -g_{b}n_b & -h_b
 \end{array}\right)
\end{equation}
where we have defined $n_{ab}=\sqrt{n_an_b}$ and 
\begin{align}
 h_a&=\frac{\hbar^2 k^2}{2m} + 2g_{a}n_a + g_{ab}n_b - \mu, \\
 h_b&=\frac{\hbar^2 k^2}{2m} + 2g_{b}n_b + g_{ab}n_a - \mu, \\
 \mu&=\frac{1}{2}\left[ g_{a}n_a + g_{b}n_b + g_{ab}n -\frac{n}{\sqrt{n_an_b}} |\Omega|\right]\,.
\end{align} 
We moreover choose the normalisation such as that the amplitudes $u_\sigma$ and $v_\sigma$ satisfy 
\begin{equation}
|u_a|^2-|v_a|^2+|u_b|^2-|v_b|^2=\pm 1\,.  \label{norm}                                                                     
\end{equation}

\subsubsection{Excitations above the symmetric GS1}\label{SecBogGS1}

\begin{figure}[h]\centering
 \epsfig{file=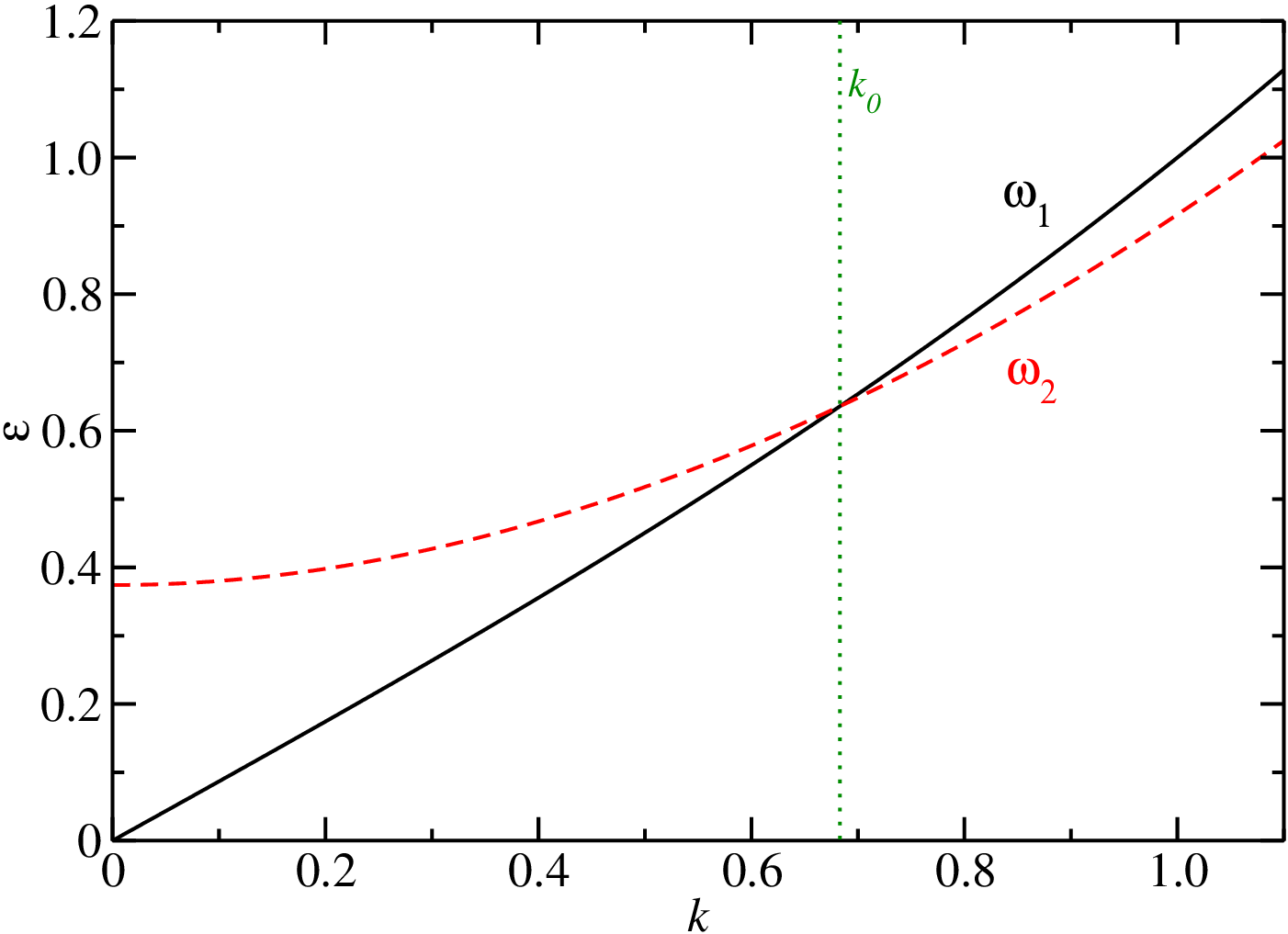,width=\linewidth,clip=true}
 \epsfig{file=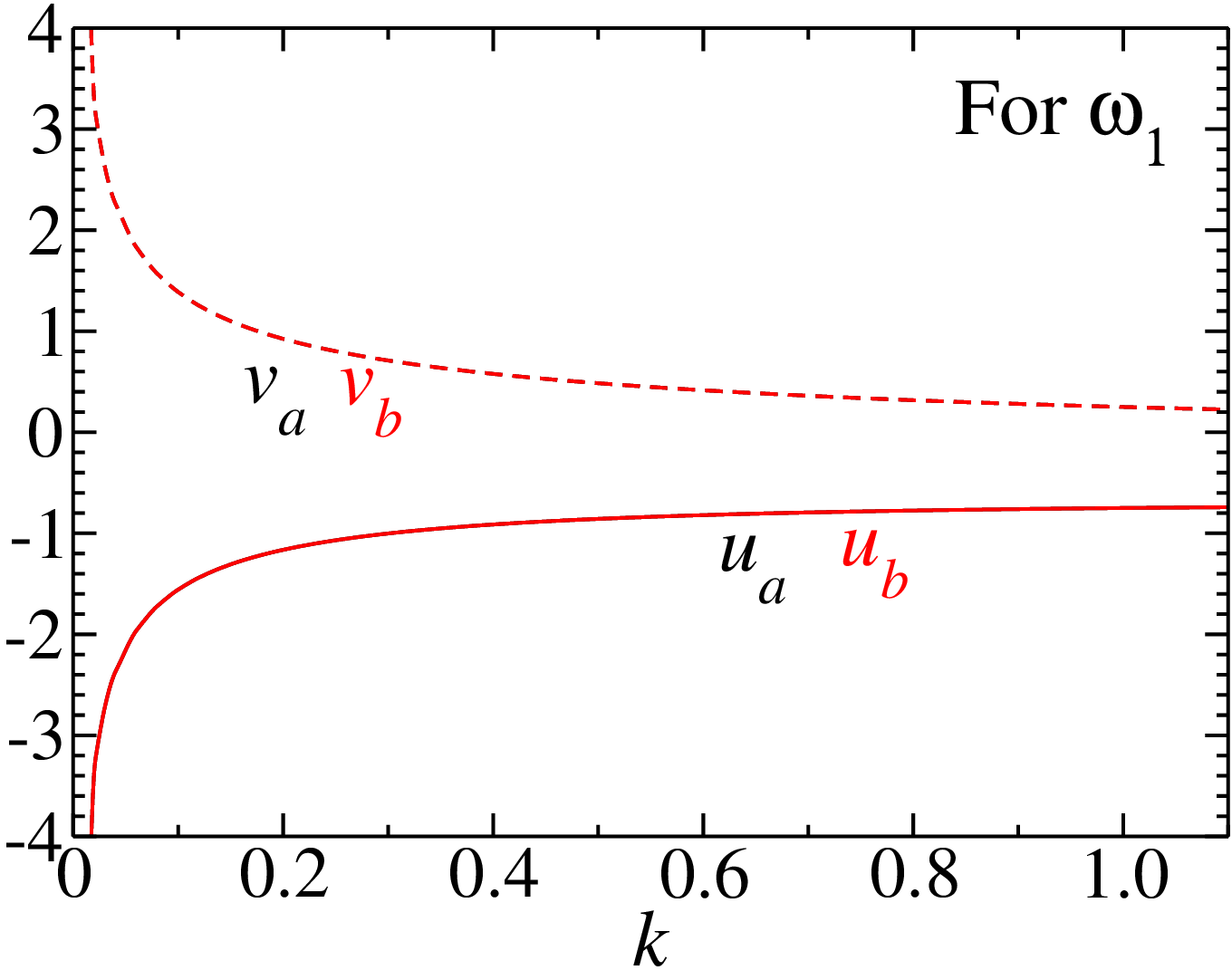,width=0.49\linewidth,clip=true}\epsfig{file=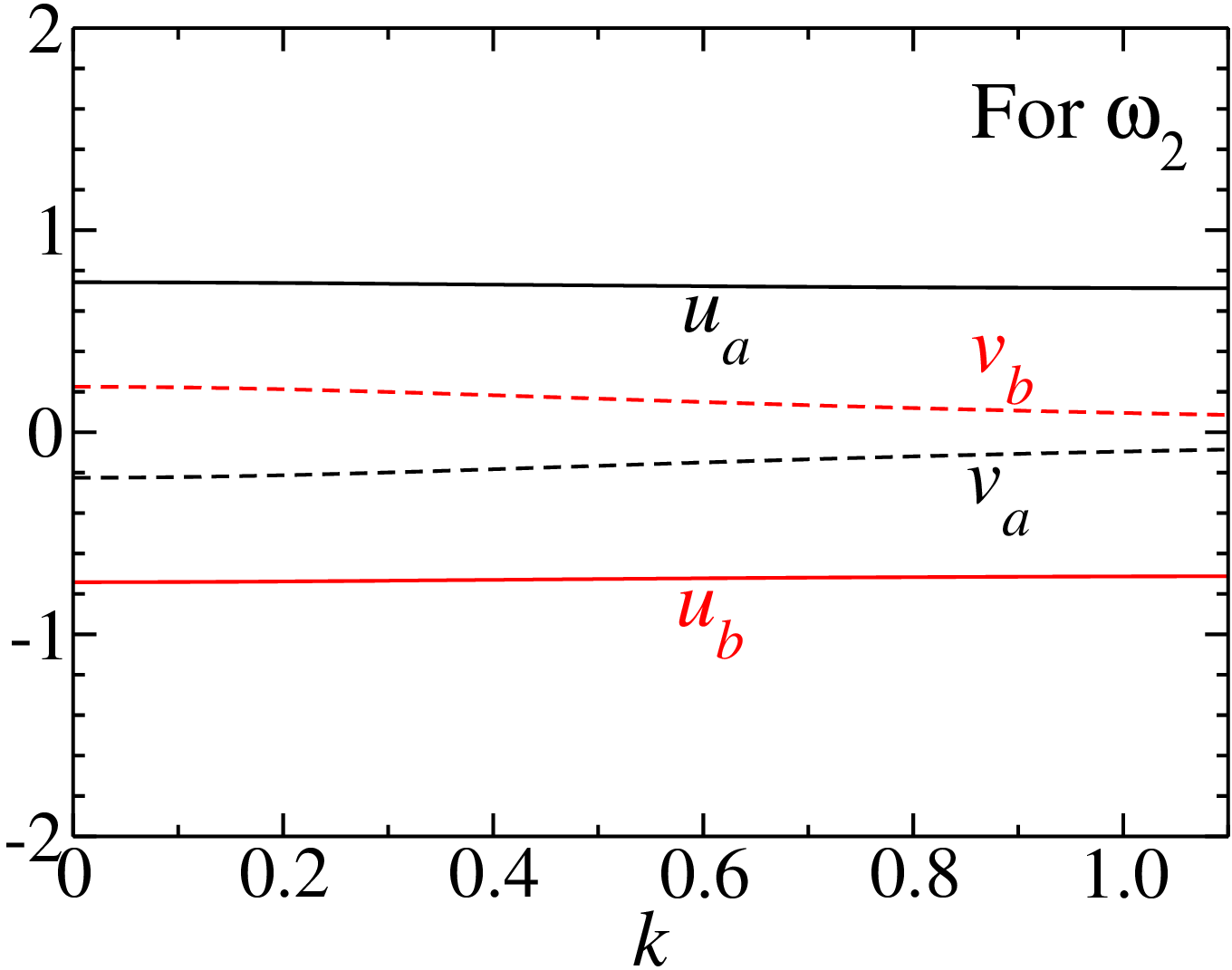,width=0.49\linewidth,clip=true}
 \caption{Top: Dispersion relation $\varepsilon(k)$ above GS1, for $g_{ab}=0.5\,g$ and $\Omega=0.1\,gn$. Bottom: Components of the eigenvectors corresponding to eigenvalues $\omega_1$ (left) and $\omega_2$ (right).}\label{figevalcasg}
\end{figure}

\begin{figure}[h]
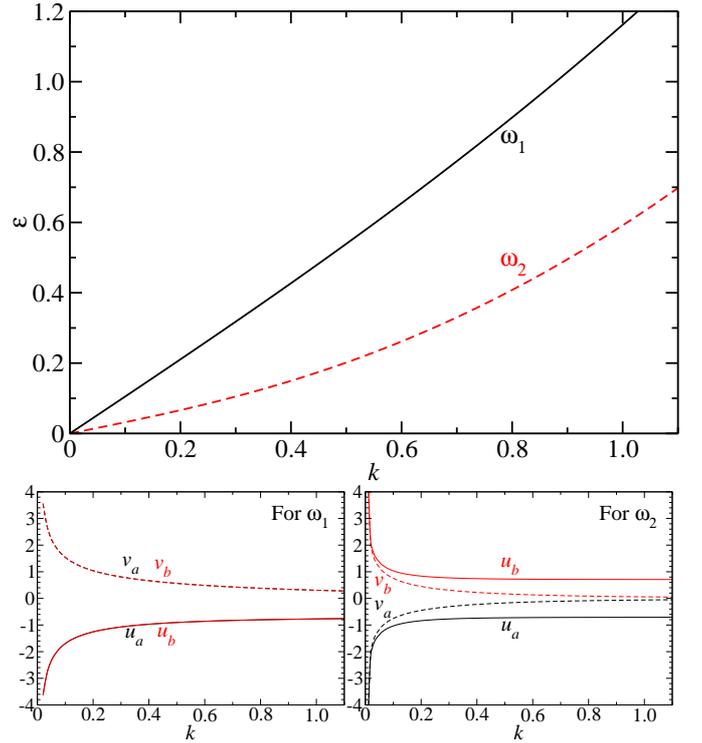
\centering
 \epsfig{file=evalcrit.eps,width=\linewidth,clip=true}
 \epsfig{file=evec4crit.eps,width=0.49\linewidth,clip=true}\epsfig{file=evec2crit.eps,width=0.49\linewidth,clip=true}
 \caption{Top: Dispersion relation $\varepsilon(k)$ at the critical point, for $g_{ab}=\,g$ and $\Omega=0.1\,gn$. Bottom: Components of the eigenvectors corresponding to eigenvalues $\omega_1$ (left) and $\omega_2$ (right).}\label{figevalcrit}
\end{figure}

Let us first analyse the excitations above GS1 for $g_a=g_b\equiv g$, which corresponds to $n_a=n_b=n/2$ (see Fig.~\ref{figgs1}). This solution only exists for $g_{ab}<g+2|\Omega|/n$. The dispersion relations are 
\begin{align}
 (\hbar\omega_1)^2=&\frac{\hbar^2k^2}{2m}\left( \frac{\hbar^2k^2}{2m} + (g +g_{ab}) n \right)\\
 (\hbar\omega_2)^2=&\frac{\hbar^2k^2}{2m}\left( \frac{\hbar^2k^2}{2m} + (g-g_{ab}) n + 4 |\Omega| \right)  +\nonumber\\
  & +2|\Omega| \big[ (g-g_{ab})n +2|\Omega| \big]
\end{align}
These results are in agreement with Refs.~\cite{Goldstein1997,Search2001,Tommasini2003} and  are plotted in Fig.~\ref{figevalcasg}. 
The frequency $\omega_1$ corresponds to a density mode as can be seen in the bottom-left panel of Fig.~\ref{figevalcasg} (see also text below). At low $k$ the dispersion is linear with sound speed $c_d$.
Notice that the frequency $\omega_1$ does not depend on the coupling $\Omega$ and is given by the same expression as the density mode of the uncoupled two-component case \cite{BECPhaseSep} (see Eq.~(\ref{mixtures}) below). 
In contrast, frequency $\omega_2$ -- which corrects the expression Eq. (\ref{wsHD}) -- corresponds to a spin mode (see bottom-right panel of Fig. \ref{figevalcasg} and text below), that goes as $\sim k^2$ for $k\to 0$ and has a gap at $k=0$ of value 
\be
  \hbar\omega_J=\sqrt{2|\Omega|\big[(g-g_{ab})n+2|\Omega|\big]} \ ,\label{omegaJ}
\ee
This gap can be shown to correspond to the Josephson frequency for small amplitude oscillations.
In the limit of $|\Omega|\to0$ we recover the result for a mixture of two components,
\begin{align}
 (\hbar\omega)^2= \frac{\hbar^2k^2}{2m}\left[ \frac{\hbar^2k^2}{2m} +  n(g \pm g_{ab})\right]. \label{mixtures}
\end{align}
As already anticipated in Sec.~\ref{SecHydro}, there is a crossing between the two frequencies at a finite value of $k$,
\be
    k_0=\sqrt{\frac{2m|\Omega|}{\hbar^2}\left(\frac{g}{g_{ab}-2|\Omega|/n}-1\right)}\,. \label{eqk0}
\ee
Notice that $k_0$ exists provided $g_{ab}<g+2|\Omega|/n$, or $g_{ab}-2|\Omega|/n<g$, which is exaclty the condition for GS1 to be the ground state.  
As the critical condition is approached the crossing occurs at lower $k$ and the gap energy approaches zero. At the critical condition the gap $\omega_J$ closes and  the dispersion relation becomes linear at low $k$, as can be seen in Fig.~\ref{figevalcrit}. Such a behavior is very different from the softening of the mode in the mixture case at the demixing point, for which the frequency goes as $\omega\sim k^2$ at low $k$. 
Finally, for $g_{ab}>\bar{g}_{ab}$ the frequency $\omega_2$ (calculated above GS1) becomes imaginary, leading to instability, since the real ground state under this condition is GS2.

To understand the density and spin characters of the modes $\omega_1$ and $\omega_2$ it is useful to look at their eigenvectors (bottom panels of Fig.~\ref{figevalcasg}). The eigenvector corresponding to $\omega_1$ is a density-like (in-phase) excitation which satisfies $u_b=u_a\equiv u_k$ and $v_b=v_a\equiv v_k$ and shows the usual infrared divergence $v_k\to k^{-1/2}$, $u_k\to -v_k$ for $k\to 0$. The eigenvector corresponding to $\omega_2$ is spin-like (out-of-phase) and satisfies $u_b=-u_a\equiv- u_k$ and $v_b=-v_a\equiv -u_k$. In this case, away from the phase transition, the eigenvector components are not strongly dependent on $k$ and they remain finite over all $k$-space at the value given by the gap. 
Approaching the phase transition the spin mode diverges for $k\to 0$ (see bottom left panel of Fig.~\ref{figevalcrit}), but $u$'s and $v$'s with the same index $a,b$ have the same sign, which will lead to a diverging spin static structure factor (see Sec.~\ref{SecSFacGS1}). It is interesting to note that there is a change of sign in $u_a$ and $u_b$ for the spin mode as soon as $g_{ab}>g$ (compare, for instance, Figs.~\ref{figevalcasg} and \ref{figevalcrit}).

\subsubsection{Excitations above the asymmetric GS1  and GS2 }\label{SecBogGS2}

\begin{figure}
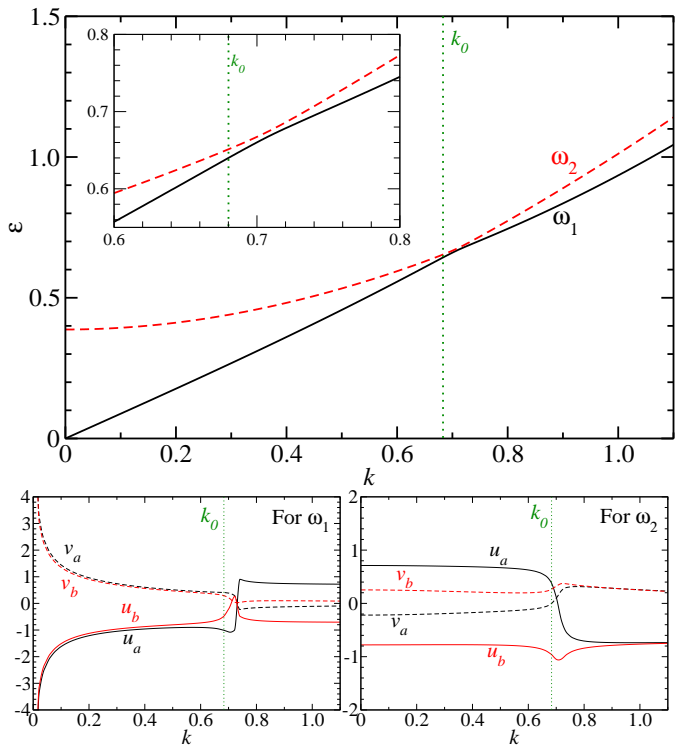
\centering
 \epsfig{file=evalgen.eps,width=\linewidth,clip=true}
 \epsfig{file=evec2gen.eps,width=0.49\linewidth,clip=true}\epsfig{file=evec4gen.eps,width=0.49\linewidth,clip=true} 
 \caption{Dispersion relation $\varepsilon(k)$ above the asymmetric GS1 ground state with $g_{ab}=0.5\, g_a$, $\Omega=0.1\, g_a n$ and
 $g_b=1.1\, g_a$. The latter induces a polarisation $(n_a-n_b)/n=-0.06$. The dotted line marks the value of the crossing $k_0$ for the corresponding case with $g_a=g_b$.  Bottom: Components of the eigenvectors corresponding to eigenvalues $\omega_1$ (left) and $\omega_2$ (right).}\label{figevalgen}
\end{figure}

From the point of view of the excitation modes, the GS2 phase and the case with $g_a\neq g_b$ are qualitatively equivalent since both situations are characterised by a finite polarisation and a coupling between density and spin-density modes (hybridization). Full analytical expressions for the frequencies $\omega_1$ and $\omega_2$ in the general case were reported in Ref.~\cite{Tommasini2003}, but here we calculate them numerically to have a clearer access to the physics. An example of the resulting avoided crossing of the modes $\omega_1$ and $\omega_2$  is given in Fig.~\ref{figevalgen} for the case $g_b\simeq g_a$. The fact that $\delta g>0$ shifts the value of $k$ at which the avoided crossing occurs at $k>k_0$. For $\delta g<0$, the avoided crossing takes place at  $k<k_0$. The bottom panels of the figure show the components of the eigenvectors corresponding to the new $\omega_1$ and $\omega_2$ modes. We can see that away from $k_0$ their values approach (perturbatively) the values of the symmetric case (Fig.~\ref{figevalcasg}), while around $k_0$ they show a sharp change of magnitude.  This change is steeper as $\delta g\to 0$, becoming a delta function for $\delta g= 0$. Notice that this fact is hidden in Fig.~\ref{figevalcasg} into the definition of the modes $\omega_1$ and $\omega_2$ before and after the crossing (compare solid and dashed lines in Figs.~\ref{figevalcasg} and \ref{figevalgen}).

\begin{figure}
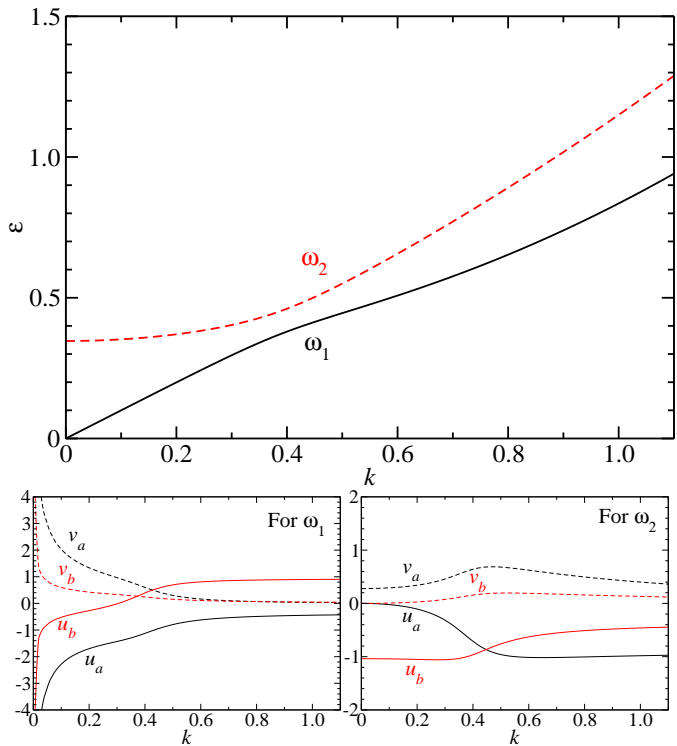
\centering
 \epsfig{file=evalgs2.eps,width=\linewidth,clip=true}
 \epsfig{file=evec2gs2.eps,width=0.49\linewidth,clip=true}\epsfig{file=evec4gs2.eps,width=0.49\linewidth,clip=true} 
 \caption{Dispersion relation $\varepsilon(k)$ above GS2 for $g_{ab}=1.4\, g$ and $\Omega=0.1\, g n$. The polarisation for this case is $(n_a-n_b)/n=-0.866$.  Bottom: Components of the eigenvectors corresponding to eigenvalues $\omega_1$ (left) and $\omega_2$ (right).}\label{figeval2gen}
\end{figure}

For the GS2 the point where the avoided crossing occurs depends strongly on the specific location along the bifurcation curve. For $n_a-n_b$ small (that is for $g_{ab}\gtrsim \bar{g}_{ab}$) the avoided crossing will show up at low $k$ and the energy difference between the two modes will be small. Instead, for a polarisation tending to unity, the avoided crossing moves to $k\to \infty$ and the energy difference between the two modes increases. We can see from the bottom panels of Fig.~\ref{figeval2gen} that when the polarisation is large (and the energy gap between the two modes at the avoided crossing grows) the behaviour of the eigenmodes departs from their behaviour in GS1.

\subsubsection{Landau critical velocity}\label{SecLandau}

One of the consequences of the spectrum discussed above is that the Landau critical velocity $v_L=\min[\omega(k)/k]$ is not just the slope of the density mode, namely $v_L\neq\lim_{k\rightarrow 0}\omega_1(k)/k=c_d$, but it is smaller. In particular for the symmetric GS1 case $v_L$ is given by
\begin{align}
 v_L=&\Bigg\{\frac{1}{m}\sqrt{2|\Omega|\left[ (g-g_{ab})n+2|\Omega| \right]}+ \nonumber\\
  &+\frac{1}{2m}\left[(g-g_{ab})n+4|\Omega|\right] \Bigg\}^{1/2}.
\end{align}
In the limit $\Omega\to0$ we recover the sound velocity of the spin mode for a mixture of two condensates, see Eq.~(\ref{mixtures}). For  $g_{ab}\to \bar{g}_{ab}$ we get $v_L\to \sqrt{(\bar{g}_{ab}-g)n/m}$. Notice that while $v_L\neq 0$ at the transition point at finite $\Omega$, for a mixture (or $\Omega=0$) $v_L=0$. 

Clearly a pure density probe would not be very effective to measure the reduction of $v_L$ since it comes mainly from spin-like  excitations. One could think to use a probe that acts only on one of the two components in such a way to be well coupled to both the density and the spin sector. Recently this topic has received some attention in a two-component system with spin-coupling in $\sigma_z$~\cite{Flayac2012}.

\section{Structure factors}\label{SecSFacGS1}

The dynamic response of a system to an external perturbation can be studied within linear response theory in terms of the dynamic structure factor
\be
   S_F(k,\omega)= \frac{1}{\hbar}\sum_{m\neq 0}\left| \Braket{m|F(k)|0}\right|^2\delta(\omega-\omega_{m0})
\ee
where $F(k)$ is the operator related to the perturbation one is interested in (see for instance Refs.~\cite{Lipparini2003,Pitaevskii2003}). 
Integration of $S_F(k,\omega)$ over the frequency $\omega$ gives the static structure factor $S_F(k)$, which quantifies how strongly the different modes are excited by the perturbation. 

In a two-component spinor condensate the interesting response functions are related to the in-phase and out-of-phase density modes, whose corresponding 
operators are 
\begin{align}
  F_d(k)&=\delta\hat{\rho}^\dag(k)=\sum_{\sigma=a,b}\left( \hat{a}^\dag_{k\sigma}\hat{a}_{0\sigma}+ \hat{a}_{0\sigma}^\dag \hat{a}_{-k\sigma} \right)\\
  F_s(k)&=\sigma_z\delta\hat{\rho}^\dag(k)=\sum_{\sigma=a,b}\text{sgn}(\sigma)\left( \hat{a}^\dag_{k\sigma}\hat{a}_{0\sigma}+ \hat{a}_{0\sigma}^\dag \hat{a}_{-k\sigma} \right)
\end{align}
where $\hat{a}^\dag_{k\sigma}$ ($\hat{a}_{k\sigma}$) creates (annihilates) a particle $\sigma=a,\;b$ in a state with momentum $k$,  $\sigma_z=\{\{1\;0\},\{ 0\; -1\}\}$ is the third Pauli matrix, and $\text{sgn}(a)=-\text{sgn}(b)=1$. 
Within the Bogoliubov approximation discussed in Sec.~\ref{SecBogExc} we can introduce the quasi-particle creation (annihilation) operators, $\hat{b}_{k\alpha}^\dag$ ($\hat{b}_{k\alpha}$) for the modes $\alpha=1,\;2$ 
\begin{align}
  \hat{a}_{k\sigma}=& e^{i\phi_\sigma}\sum_{\alpha=1,2}(u_{k\sigma\alpha}\hat{b}_{k\alpha} + v_{-k\sigma\alpha}^* \hat{b}_{-k\alpha}^\dag) \label{Bogquant1}\\
  \hat{a}_{k\sigma}^\dag=& e^{-i\phi_\sigma}\sum_{\alpha=1,2}(u_{k\sigma\alpha}^*\hat{b}_{k\alpha}^\dag + v_{-k\sigma\alpha} \hat{b}_{-k\alpha}) \label{Bogquant2}
\end{align}
where $\{u_{k\sigma\alpha},\;v_{k\sigma\alpha}\}$ are the eigenvectors related to the $\alpha$ mode found in Sec.~\ref{SecBogExc}.
The density and spin dynamic structure factors are easily calculated and can be written, respectively, as
\begin{align}
  S_d(k,\omega)=&S_d^1(k)\delta(\omega-\omega_1) + S_d^2(k)\delta(\omega-\omega_2), \\
  S_s(k,\omega)=&S_s^1(k)\delta(\omega-\omega_1) + S_s^2(k)\delta(\omega-\omega_2),
\end{align}
with
\begin{align}
   S_d^\alpha(k)&=\frac{1}{\hbar}\left|\sum_\sigma \sqrt{n_\sigma}\left( u_{k\sigma\alpha}+v_{k\sigma\alpha}\right)\right|^2,\label{Sden}\\
   S_s^\alpha(k)&=\frac{1}{\hbar}\left|\sum_\sigma \sqrt{n_\sigma}\,\text{sgn}(\sigma)\left( u_{k\sigma\alpha}+v_{k\sigma\alpha}\right)\right|^2. \label{Sspin}
\end{align}
We immediately see that for the symmetric GS1, due to the symmetry of Bogoliubov amplitudes $u$ and $v$, the density structure factor is completely exhausted by the density mode $\omega_1$, while the spin structure factor is completely exhausted by the spin mode $\omega_2$. 
On the contrary both for GS2 and the asymmetric case (i.e., $g_a\neq g_b$), the interference term in the equations becomes important due to mode hybridization.

\begin{figure}
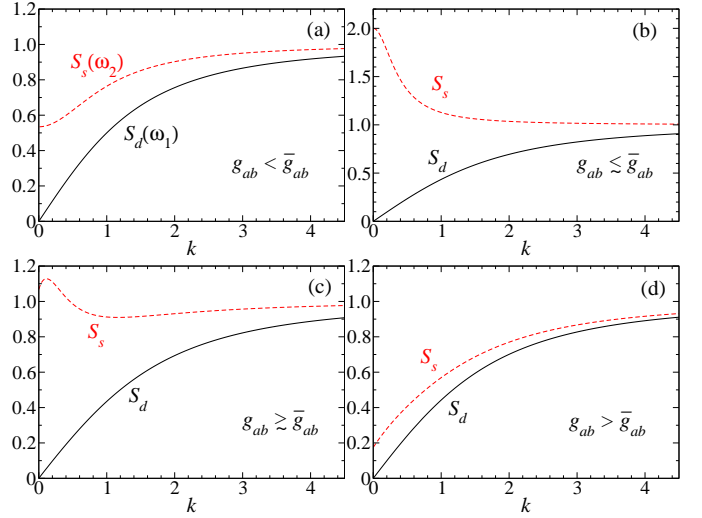

 \epsfig{file=sfactor.eps, width=0.5\linewidth, clip=true}%
 \epsfig{file=sfacgs1limk0.eps, width=0.5\linewidth, clip=true}\\
 \epsfig{file=sfaccrit.eps, width=0.5\linewidth, clip=true}%
 \epsfig{file=sfacgs2.eps, width=0.5\linewidth, clip=true} 
 \caption{Density (continuous black line) and spin (red dashed line) static structure factors for $g_a=g_b$ for GS1 with (a) $g_{ab}/g_a=0.5$, (b) $g_{ab}/g_a=1.15$ and for GS2 with (c) $g_{ab}/g_a=1.25$, (d) $g_{ab}/g_a=1.5$. 
 In all cases, $\Omega=0.1 g_a n$ and $\bar{g}_{ab}/g_a=1.2$.}\label{sfactor}
\end{figure}

The static structure factors are simply \be  S_{d(s)}(k)=S^1_{d(s)}(k) + S^2_{d(s)}(k)\ee
and are plotted for $g_a=g_b$ in Fig.~\ref{sfactor} as a function of $k$, for four different values of $g_{ab}$ before and after the transition. 
As expected the density structure factor shows the usual linear behaviour at low $k$ found for a condensate, and it satisfies Feynman criterion (see for instance Ref.~\cite{Pitaevskii2003}). 

The spin structure factor, instead, shows a gap at $k=0$ which arises from single-particle coherence (see also next section). 
This behaviour of the spin structure factor is fundamentally different to what is usually found in Bose gases at $T=0$, and in particular in mixtures of two components, where the spin structure factor fulfills Feynman rule (see for instance \cite{Sun2010}).
At $g=g_{ab}$ the structure factor reaches unity and keeps constant for all $k$. This is related to the change of sign in $u_a$ and $u_b$ for $g=g_{ab}$ (see Sec.~\ref{SecBogGS1}). 
Approaching the phase transition, Fig.~\ref{sfactor}(b), the spin structure factor presents a maximum for $k\to 0$, which diverges at precisely $g_{ab}=\bar{g}_{ab}$. This is due to the fact that close to the phase transition spin fluctuations are very large and, since they are proportional to the spin static structure factor (see next section), this latter is also very large. Above the transition point, Fig.~\ref{sfactor}(c), the spin structure factor maximum moves toward $k\neq 0$ due to the polarisation of GS2. Finally when the system lies deep in the GS2,  Fig.~\ref{sfactor} (d), the spin-mode reduces essentially to a single-component density mode and for this reason behaves linearly at small $k$ and the gap tends to zero. 

The effect of finite temperature $T$ well below the critical condensation temperature  $T_c$ can be introduced by using detailed balanced. The static structure factors read
\be
  S_\nu(k,T)=\sum_{\alpha=1,2} S_\nu^\alpha(k) \coth\frac{\hbar\omega_\alpha(k)}{2k_B T},	\;\;\;\nu=d,s.
\ee
The zero-temperature results are recovered for $k_BT\ll\hbar\omega_\alpha(k)$. For large $\Omega$ (deep in GS1) the latter condition can be easily satisfied for the spin-mode. This is not true close to phase transition. On the other side of the phase transition the maximum at finite $k$ will be hidden by the usual enhancement of the static structure factor for $k\to 0$. 

\subsection{Density and polarisation fluctuations}
\label{Secfluc}

The peculiar behavior of the structure factors will show up in the local density $\Delta N^2$ and polarisation $\Delta M^2$ fluctuations of the spinor gas.
Such fluctuations, which are nowadays experimentally accessible (see e.g., \cite{EsslingerFluc,KetterleFluc,KetterleSpin,ChinFluc,ChinSq,1Dfluc}), are important quantities characterising a system (for a recent review in the context of cold gases see \cite{KlawunnFluc}).  
If $R$ is the linear size of the spot where the fluctuations are measured one can approximate the fluctuations as
\begin{eqnarray}
\Delta N^2 &=& n\!\int S_d(\mathbf{k},T) H(\mathbf{k})\frac{d\mathbf{k}}{(2\pi)^D}\simeq N S_d(1/R,T),\\
\Delta M^2 &=& n\!\int S_s(\mathbf{k},T) H(\mathbf{k})\frac{d\mathbf{k}}{(2\pi)^D}\simeq N S_s(1/R,T);
\end{eqnarray}
with $D$ the dimensionality of the system and $H(\mathbf{k})$ the Fourier transform of a geometrical factor that depends on the shape of the probe cell where the measurement is performed. The last approximate equality \cite{KlawunnFluc}, where $R$ is the linear size of the spot, has been recently used to probe the (density) structure factor of a 2-dimensional Bose gas \cite{ChinSq}.
For a spinor gas the local fluctations of the polarisation can be also used to measure the spin structure factor.

The behaviour of $S_s(0)$ can be easily understood by looking at the spin fluctuations. Indeed, if we look at them at the single-particle level (two level system), we always find that $\Delta M^2=\langle \sigma_z^2\rangle-\langle \sigma_z\rangle^2\neq0$. Since we have the relation $S_s(0)=\Delta M^2$ it follows that the spin structure factor is gapped. In this sense we can say that the gap in $S_s$ comes from single-particle coherence.

On the other hand since the phase transition is due to spin-density instabilities, as we have already seen from the susceptibility, Sec.~\ref{SecGS}, close to the phase transition the fluctuations of the polarisation, i.e., $S_s(0)$, diverge. For instance, in the context of Bose-Bose mixtures such fluctuation enhancement has been very recently observed in a quenching experiment \cite{De2012}.

\section{Trapped two-component spinor condensates}\label{SecTrap}

In this section we discuss how the GS1--GS2 picture presented in Sec.~\ref{SecGS} can be applied to inhomogeneous systems. 
We find that local density approximation captures the main features of the system brought about by the trapping potentials. Within this approach the equations describing the ground state of the trapped system can be found from Eqs.~(\ref{tdgppsia}) and (\ref{tdgppsib}). Neglecting the kinetic energy we find by addition and subtraction the two equations
\begin{eqnarray}
  \!\!\!\!\!\left(g-g_{ab}+\frac{|\Omega|}{\sqrt{n_an_b}}\right)(n_a-n_b)&=&V_b-V_a, \label{GSnonhom}\\
  \!\!\!\!\!\left(g+g_{ab}-\frac{|\Omega|}{\sqrt{n_an_b}}\right)(n_a+n_b)&=&2\mu-(V_b+V_a). \label{GSnonhom2}
\end{eqnarray}
For $V_a=V_b=0$ we obviously recover the results from Sec.~\ref{SecGS}.
To exemplify what happens for $V_a, V_b\neq 0$, we consider two situations. In the first one, the trapping is harmonic and equal for species $a$ and $b$. In the second example, instead, one of the components is subject to an optical lattice potential, which is not directly seen by the second component. In both cases, numerical solutions of the full GP equations are provided for comparison.

\subsection{Harmonically trapped two-component spinor}
\label{SecHTrap}

Let us consider what happens if there is a spherically symmetric external harmonic confinement $V_{\text{ho}}(r)=m\omega^2 r^2/2$, which acts in the same way on both spinor components. We first address this problem within local density approximation. For $V_a=V_b$ the right-hand side of Eq.~(\ref{GSnonhom}) vanishes, which means that there are two possible ground states characterised locally by the same solutions as the homogeneous system, Eqs.~(\ref{nsym}) and (\ref{npm}).
This allows us to introduce the local critical value $\bar{g}_{ab}(r)=g+2|\Omega|/n(r)$, with $n(r)$ the total local density. Analogously to the homogeneous case, for $\bar{g}_{ab}(r)>g_{ab}$ the system is locally in GS1, while for the opposite condition the system in locally in GS2. 
Since at low density the Rabi flopping term always dominates, when a two-component spinor condensate is in a trap two possible scenarios can exist:

(i) the whole system is in GS1, i.e.,  the critical condition $g_{ab}<\bar{g}_{ab}(0)$ is fulfilled; 

(ii) GS2 is the lowest energy state in the center of the trap, i.e., $g_{ab}>\bar{g}_{ab}(0)$. Then there always exsists some critical radius $R_c$ fulfilling that the (decreasing) density is such that $g_{ab}<\bar{g}_{ab}(r)$ for $r>R_c$ and GS1 is the lowest energy state. In this case there is coexistence of the two phases and the critical radius is given by $g_{ab}=\bar{g}_{ab}(R_c)$. 

\begin{figure}\centering
   \epsfig{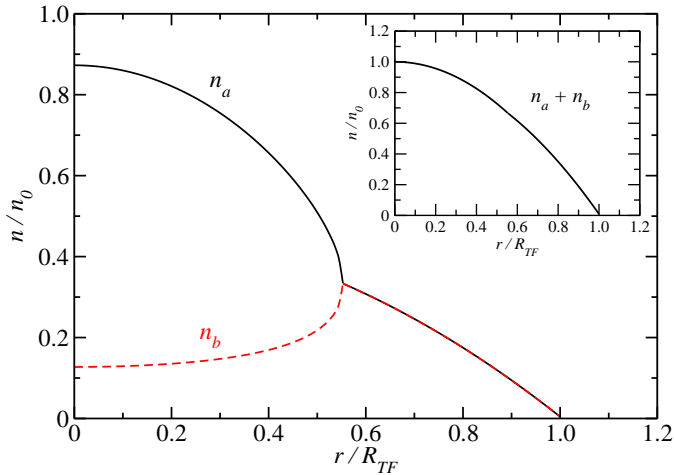}
  \caption{Example of the ground state in a harmonic trap, with an inner region described by GS2 and an outer region described by GS1. Parameters: $g_{ab}/g=1.3$, $\Omega/\mu=0.1$ and $R_c/R_{TF}\simeq 0.55$. The quantities $n_0$ and $R_{TF}$ represent, respectively, the density at $r=0$ and the radius where the density vanishes.}\label{figtrap}
\end{figure}

The first scenario is similar to the usual Thomas-Fermi approximation for a single condensate. We concentrate therefore on the second situation.
For large $r$ the system will be in GS1, that is, $n_a(r)=n_b(r)=n(r)/2$. For a spherically symmetric harmonic potential the density in the GS1 phase can be calculated from Eq.~(\ref{GSnonhom2}) as
\begin{equation}
 n(r)=2\frac{\mu - V_\text{ho}(r)}{g+g_{ab}-2|\Omega|/n(r)}
\end{equation}
At the critical radius the density fulfills $n(R_c)=2|\Omega|/(g_{ab}-g)$ and substituting above we find
\begin{equation}
R_c=\sqrt{2\mu+4|\Omega|\frac{ g}{g-g_{ab}}},\;\;\;\;g>g_{ab}. 
\end{equation}
For $r>R_c$ the system is in GS1, while for $r<R_c$ GS2 is the ground state and the system is polarised. Figure~\ref{figtrap} shows the density distribution of components $a$ and $b$, as well as the total density (in the inset), of a 2-phase configuration. Due to the use of the local density approach the density profiles measurement is a direct mapping of the phase diagram of the homogeneous system, Fig.~\ref{figgs1}.

\begin{figure}
   \epsfig{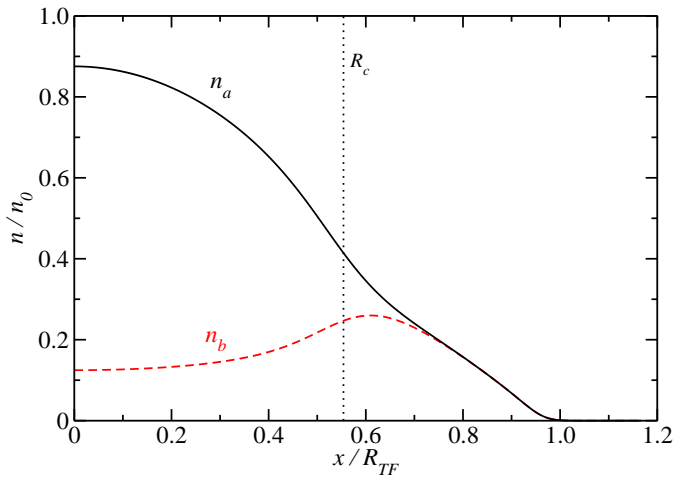}
 \caption{Two-phase ground state density profiles of components $a$ and $b$ from the numerical solution of one-dimensional GP equation. The parameters are: $g_{ab}/g=1.25$, $\Omega/\mu=0.05$ and $R_c/R_{TF}=0.55$. The quantities $n_0$ and $R_{TF}$ represent, respectively, the density at $x=0$ and the position at which the density vanishes. }\label{figtrapgp}
\end{figure}

Including quantum pressure the density profiles are smoothed out. This behavior is shown in Fig.~\ref{figtrapgp}, obtained by numerically solving the quasi-one-dimensional spinorial GP equations in a trap. We have checked that equivalent density distributions are found in the 3D case for a spherically symmetric trap.

The two-phase configuration that we find is reminiscent of the results reported in Ref.~\cite{Merhasin2005} in the intermediate state in the transition from the miscible to immiscible regimes (see also the experiment of Ref.~\cite{Nicklas2011}).

\subsection{Spinor condensate in a spin-selective optical lattice}
\label{SecOL}

Interesting scenarios appear when the two components feel different external potentials. We concentrate on the situation where component $a$ feels an optical lattice potential, while component $b$ does not. Since the atomic transition is locally out-of-resonance, polarised ground states are favoured, as can be seen from Eq.~(\ref{GSnonhom}). 
Assuming $|\delta V|=|V_a-V_b|\ll |V_a+V_b|$ we can perform a perturbative analysis. We define $V_\sigma=V_0+\delta V_\sigma$, with $\sigma=a,b$, which allows us to write the densities as $n_\sigma=n_\sigma^0+\delta n_\sigma$, where the $n_\sigma^0$ are the unperturbed densities, that is the densities corresponding to $V_0$. 
For GS1 $n_a^0(x)=n_b^0(x)=n_0(x)/2$ and we find at first order
\be
 \delta n_{\sigma}=\frac{\delta\mu - \Delta V/2}{g+g_{ab}}-\text{sgn}(\sigma)\frac{\delta V/2}{2|\Omega|/n_0+g-g_{ab}},\label{deltanol}
\ee
where $\delta\mu$ is the change in chemical potential and $\Delta V=\delta V_a + \delta V_b$.
Notice that from this expression the external potential acts on component $a$ but also on component $b$. This means that the interspecies interaction as well as the Rabi coupling create an effective external potential for the untrapped component.

The first term in the right-hand side of Eq.~(\ref{deltanol}) acts as renormalized chemical potential that takes into account a mean potential change, $\Delta V/2$, and gives the same contribution to the densities. The second term instead is species-dependent, showing a positive or negative contribution depending on the component.
When the first term dominates, the fluctuations $\delta n_\sigma$ have the same sign, while when the second term dominates $\delta n_a$ and $\delta n_b$ are out-of-phase. It is clear then that for large $\Omega$ or equivalently small densities the system tends to be unpolarised; in contrast, for high densities (for instance, at the trap center) the system tends to be polarised.
 
Obviously the density change in component $a$ is opposite to $\delta V$ and close to the phase transition even a small difference in the external potential gives rise to a large local polarisation, and eventually the perturbative approach breaks down.
A similar solution and argument hold for GS2.

\begin{figure}\centering
   \epsfig{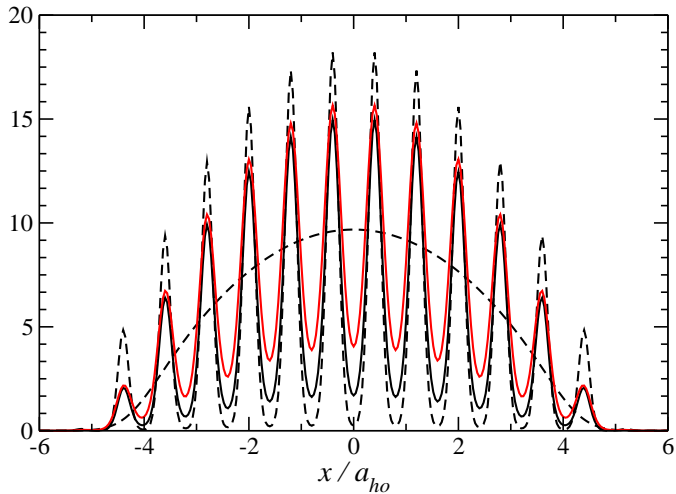}
  \caption{Density profiles (solid line) for a spinor condensate in a spin-dependent optical lattice plus harmonic trap (see text) in GS1. The parameters are $g_{ab}=0$, $\Omega/(gn)=1$, $V_0/(gn)=1$ and $d=0.8\,a_{ho}$, with $a_{ho}=\sqrt{\hbar/m\omega}$ the oscillator length. Dark line shows $n_a$ and light (red) line shows $n_b$. We report for comparison also the case with $\Omega=0$ (dashed lines). }\label{figgs1ol}
\end{figure}

\begin{figure}\centering
   \epsfig{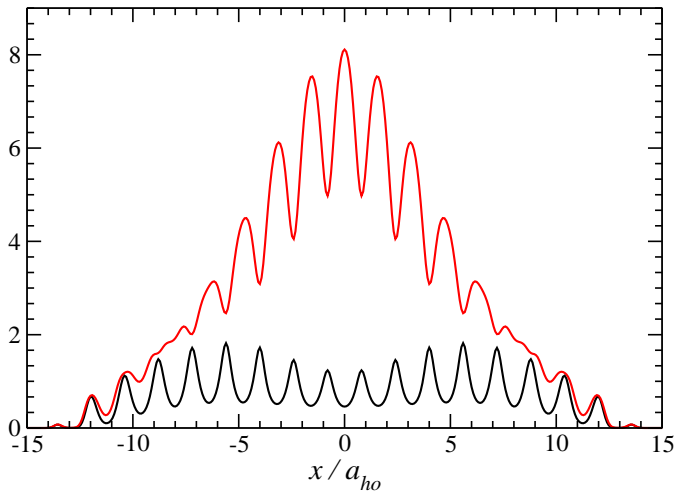}
  \caption{The same as in Fig. \ref{figgs1ol}, but in coexistence case of both GS1 and GS2. The parameters are : $g_{ab}/g=40$, $V_0/(gn)=1$, $\Omega/(gn)=1$ and $d=1.6\,a_{ho}$, with $a_{ho}=\sqrt{\hbar/m\omega}$ the oscillator length. Dark (black) solid line shows $n_a$ and light (red) solid line shows $n_b$.}\label{figgs2ol}
\end{figure}

To test the predictions of local density approximation, we numerically find the stationary ground state of the (quasi-one-dimensional) spinor Gross-Pitaevskii Eqs. (\ref{tdgppsia})-(\ref{tdgppsib}) for potentials $V_a=V_{\text{ho}} + V_{\text{ol}}$ and $V_b=V_{\text{ho}}$, with $V_{\text{ol}}(x)=V_0\left(\sin^2(k_L x)-1/2\right)$ and $V_{\text{ho}}(x)=m\omega^2 x^2/2$. In this case it is clear that  for small $g_{ab}$ the system is everywhere in the state GS1 and the effect of the Rabi coupling is just to make the state $b$ to feel the lattice, obtaining a polarised in-phase periodic spinor condensate, as the one shown in Fig. \ref{figgs1ol}. On the other hand if $g_{ab}$ is large enough we have a central region where the system is in the GS2 state, which results in a polarised out-of-phase spinor condensate as shown in Fig. \ref{figgs2ol}. As discussed above when the density is small the system enters in GS1 and this is clearly seen in the tails of the condensate, where the oscillations of the two components are in-phase.  

\section{Discussion and conclusions}

In this article we have addressed the properties of two-component condensates in the presence of a Rabi coupling. This system is at the basis of some of the current hot research areas, such as physics of spin-orbit coupled condensates. We have provided a thorough discussion of the properties of the system, recovering the results already present in the literature and adding new results and a complete description and interpretation of the fundamental properties of the system.

In light of our results, we can compare a 2-component condensate with and without the Rabi coupling term $\Omega\sigma_x$.
In the former case the transition is second order leading to a partial global polarization, while in the latter case it is first order and coexistence of domains of polarization $\pm 1$ are present in the system.
The spectrum shows a gap for $\Omega\neq 0$ since it breaks the $O(2)$ symmetry transverse to the polarization,
making excitation about $\sigma_x$ ($\sigma_y$) massive (Fig. \ref{figevalcasg}). The second order phase transition is due to the closing of the spin gap when the inter-species interaction is large enough (see Fig. \ref{figevalcrit}). For $\Omega=0$ the spectrum is gapless since any angle in the $\sigma_x$-$\sigma_y$ plane (of the Bloch sphere) is equivalent. In this case the phase transition is driven by the softening of the phonon-like spin-mode.

The excitation spectrum behaviour is reflected in the density and spin static structure factors. The first one has the same behaviour as in a single component condensate and satisfies the Feynman relation. The spin structure factor instead shows a peculiar behaviour and in general does not follow the Feynman relation. In particular due to the coherent coupling it remains finite at zero momentum, and it shows a divergence approaching the phase transition (Fig. \ref{sfactor}). Such a divergence is easily explained in terms of the fluctuations of the polarization (see Sec. \ref{Secfluc}).

We also discuss the coexistence of the two ground states once the gas is confined in a harmonic potential.
In this case the phase diagram can be almost directly read out by measuring the local density of the two components, as clearly seen in Fig. \ref{figtrap}.

Even more interesting is the effect of a spin-dependent external potential, which creates an entanglement between the internal and the external degrees of freedom of the gas. The possible configurations are many and we concentrate on the case of an optical lattice potential $V_{\text OL}$ felt by one (component $a$) of the two components, i.e., a potential $V_{\text OL}\otimes(1+\sigma_z)$. The results obtained numerically solving coupled GP equations are reported in Figs. \ref{figgs1ol}-\ref{figgs2ol}, where we show how depending on the regime the periodic structure induced on the ``free'' component $b$ can be either in- or out-of-phase with respect to the trapped component $a$.

\vspace{1em}
We acknowledge very useful discussions with I. Carusotto, Y. Li, D. Papoular, S. Stringari and A. Trombettoni. This work has been financially supported by ERC through the QGBE grant and by Provincia Autonoma di Trento.


\thebibliography{99}

\bibitem{andreev}
S. I. Shevchenko, and D. V. Fil, JETP {\bf 105}, 135 (2007).
 
\bibitem{ZoranSC} S. Beattie, S. Moulder, R. J. Fletcher, and Z. Hadzibabic, Phys. Rev. Lett. {\bf 110}, 025301 (2013).

\bibitem{ChapmanJJ}
M.-S. Chang, Q. Qin, W. Zhang, L. You and M. S. Chapman, Nature Physics {\bf 1}, 111 (2005).

\bibitem{Zibold2010} 
T. Zibold, E. Nicklas, C. Gross, and M. K. Oberthaler, Phys. Rev. Lett. {\bf 105}, 204101 (2010).

\bibitem{Cirac1998} 
J. I. Cirac, M. Lewenstein, K. Molmer, and P. Zoller, Phys. Rev. A {\bf 57}, 1208 (1998).

\bibitem{TwinExp}
C. Gross, H. Strobel, E. Nicklas, T. Zibold, N. Bar-Gill, G. Kurizki, and M. K. Oberthaler,
Nature {\bf 480}, 219 (2011);
B. L\"ucke, M. Scherer, J. Kruse, L. Pezz\'e, F. Deuretzbacher, P. Hyllus, O. Topic, J. Peise,
W. Ertmer, J. Arlt, L. Santos, A. Smerzi, C. Klempt, Science {\bf 334}, 773 (2011).

\bibitem{Uwe} U. R. Fischer and R. Sch\"utzhold, Phys. Rev. A {\bf 70}, 063615 (2004).

\bibitem{Liberati2006} S. Liberati, M. Visser, and S. Weinfurtner, Class. Quant. Grav. {\bf 23}, 3129 (2006).

\bibitem{Sindoni2011} L. Sindoni, Phys. Rev. D {\bf 83}, 023011 (2011).

\bibitem{texture}
K. Kasamatsu, M. Tsubota, and M. Ueda, Int. J. of Mod. Phys. B 19, 1835 (2005).

\bibitem{DalibardRMP}
J. Dalibard, F. Gerbier, G. Juzeliunas, A. Gostauto, P. Ohberg, Rev. Mod. Phys. {\bf 83}, 1523 (2011).




\bibitem{Goldstein1997} E. V. Goldstein and P. Mestre, Phys. Rev. A {\bf 55}, 2935 (1997).

\bibitem{Search2001} C.P. Search, A. G. Rojo, and P. R. Berman, Phys. Rev. A {\bf 64},013615 (2001).

\bibitem{Tommasini2003} P. Tommasini, E. J. V. de Passos, A. F. R. de Toledo Piza, M. S. Hussein, and E. Timmermans, Phys. Rev. A {\bf 67}, 023606 (2003).

\bibitem{Lee2004}
C. Lee, W. Hai, L. Shi and K. Gao, Phys. Rev. A {\bf 69}, 033611 (2004).

\bibitem{Liang2010} C. Liang, K. Wei, B. J. Ye, H. M. Wen, X. Y. Zhou, and R. D. Han, J. Low Temp. Phys. {\bf 161}, 334 (2010).

\bibitem{Cornell99} M. R. Matthews, B. P. Anderson, P. C. Haljan, D. S. Hall, M. J. Holland, J. E. Williams, C. E. Wieman, and E. A. Cornell, Phys. Rev. Lett. {\bf 83}, 3358 (1999).

\bibitem{Blakie1999} P. B. Blakie, R. J. Ballagh, and C. W. Gardiner, J. Opt. B: Quantum Semiclass. Opt. {\bf 1}, 378 (1999).

\bibitem{Search2001b} C. P. Search and P. R. Berman, Phys. Rev. A {\bf 63}, 043612 (2001).

\bibitem{Jenkins2003} S. D. Jenkins and T. A. B. Kennedy, Phys. Rev. A {\bf 68}, 053607 (2003).

\bibitem{Nicklas2011} E. Nicklas, H. Strobel, T. Zibold, C. Gross, B. A. Malomed, P. G. Kevrekidis, and M. K. Oberthaler, Phys. Rev. Lett. {\bf 107}, 193001 (2011).

\bibitem{Leggett2001} A. J. Leggett, Rev. Mod. Phys. {\bf 73}, 307 (2001).

\bibitem{Recati2003}
A. Recati, P. O. Fedichev, W. Zwerger, I. von Delft, P. Zoller, Phys. Rev. Lett. {\bf 94}, 040404 (2005).

\bibitem{Daley2008} A. J. Daley, M. M. Boyd, J. Ye, and P. Zoller, Phys. Rev. Lett. {\bf 101}, 170504 (2008).

\bibitem{BECPhaseSep}
{\it Bose-Einstein Condensation in Dilute Gases}, by C.~J.~Pethick and H.~Smith (Cambridge University Press, 2002).

\bibitem{Popov}
{\it Functional Integrals and Collective Excitations}, by N. V. Popov, (Cambridge University Press, 1987).

\bibitem{Flayac2012} H. Flayac, H. Ter\c cas, D. D. Solnyshkov, and G. Malpuech, arXiv: 1212.5894 (2012).

\bibitem{Lipparini2003} 
{\it Modern Many-Particle Physics. Atomic Gases, Quantum Dots and Quantum Fluids}, by E. Lipparini (World Scientific, 2003). 

\bibitem{Pitaevskii2003} {\it Bose-Einstein Condensation}, by L. Pitaevskii and S. Stringari (Oxford Science Publications, 2003).

\bibitem{Sun2010} B. Sun and M. S. Pindzola, J. Phys. B: At. Mol. Opt. Phys. {\bf 43}, 055301 (2010).

\bibitem{EsslingerFluc}
T. M\"uller, B. Zimmermann, J. Meineke, J.-P. Brantut, T. Esslinger, and H. Moritz, Phys. Rev. Lett. {\bf 105}, 040401 (2010). 

\bibitem{KetterleFluc}
C. Sanner, E. J. Su, A. Keshet, R. Gommers, Y. I. Shin, W. Huang, and W. Ketterle, Phys. Rev. Lett. {\bf 105}, 040402 (2010). 

\bibitem{KetterleSpin}
C. Sanner, E. J. Su, A. Keshet, W. Huang, J. Gillen, R. Gommers, and W. Ketterle, Phys. Rev. Lett. {\bf 106}, 010402 (2011). 

\bibitem{ChinFluc}
C.-L. Hung, X. Zhang, N. Gemelke, and C. Chin, Nature {\bf 470}, 236 (2011).

\bibitem{ChinSq}
C.-L. Hung, X. Zhang, L.-C. Ha, S.-K. Tung, N. Gemelke, and C. Chin, New Journal of Physics {\bf 13}, 075019 (2011).

\bibitem{1Dfluc}
J. Armijo, T. Jacqmin, K. V. Kheruntsyan, and I. Bouchoule,
Phys. Rev. Lett. {\bf 105}, 230402 (2010); J. Armijo, T. Berrada, K. V. Kheruntsyan, and I. Bouchoule, ibid. 106, 230405 (2011);
J. Armijo, Phys. Rev. Lett. {\bf 108}, 225306 (2012).

\bibitem{KlawunnFluc}
M. Klawunn, A. Recati, L. P. Pitaevskii, S. Stringari, Phys Rev. A {\bf 84}, 033612 (2011).

\bibitem{De2012} S. De, D. L. Campbell, R. M. Price, A. Putra, B. M. Anderson, and I. B. Spielman, arXiv:1211.3127v1 (2012).

\bibitem{Merhasin2005} I. M. Merhasin, B. A. Malomed, and R. Driben, J. Phys. B: At. Mol. Opt. Phys. {\bf 38}, 877 (2005).
\end{document}